\documentstyle[aps,twocolumn,graphicx]{revtex}

\input{tcilatex}

\begin{document}
\title{Evidence of a glass transition induced by rigidity self-organization in a
network forming fluid.}
\author{Adri\'{a}n Huerta, Gerardo G. Naumis}
\address{$^{1}$Instituto de Fisica, Universidad Nacional Aut\'{o}noma de\\
M\'{e}xico (UNAM)\\
Apartado Postal 20-364, 01000, Distrito Federal, Mexico.}
\date{\today }
\maketitle
\pacs{64.70.Pf, 64.60.-i, 05.70.-a, 05.65.+b}

\abstract{A Monte Carlo method is used in order to simulate the
competition between the molecular relaxation and crystallization times
in the formation of a glass.
The results show that nucleation is avoided during supercooling and
produce self-organization in the sense of the rigidity theory, 
where the number of geometrical constraints due to bonding and excluded volume 
are compared with the degress of freedom available to the system.
Following this idea, glass transitions were obtained by producing
self-organization, and in the case of geometrical frustration,
self-organization is naturally observed.}

\section{Introduction}

When a liquid melt is cooled, usually two things can happen: the melt
crystallize, or if the speed of cooling is high enough to avoid equilibrium,
a solid without long range order (a glass) is formed. This last process is
known as the glass transition (GT), and although is very important from the
fundamental and technological point of view, there are still many unsolved
questions related to it \cite{Anderson}. Not all materials are able to form
glasses, and many criteria have been proposed in order to explain the
ability of a material to reach the glassy state \cite{Jackle}. \ The
temperature where the GT occurs, is called the glass transition temperature (%
$T_{g}$). Many factors determine the $T_{g}$, but among these, the chemical
composition is fundamental. Chalcogenide glasses (formed with elements of
the VI column) are very useful for understanding the effects of the chemical
composition \cite{Boolchand1}. In fact, $T_{g}$ can be raised or lowered by
adding impurities, and the fragility of the glass can be changed from strong
to fragile \cite{Tatsumisago}. Recently, by using stochastic matrices \cite
{Kerner1,Kerner2}, the empirical modified Gibbs-DiMarzio law that accounts
for the relation between $T_{g}$ and the concentration of modifiers \cite
{Sreeram} has been obtained, including the characteristic constant that
appears in the law for almost any chalcogenide glass \cite{Micoulaut}.

The ease of glass formation in covalent glasses can be explained at least in
a qualitative way by the rigidity theory (RT), introduced first by Phillips 
\cite{Phillips1} and further refined by Thorpe \cite{Thorpe0}. By
considering the covalent bonding as a mechanical constraint, within this
theory, the ease of glass formation is related with the proportion of
available degrees of freedom and the number of constraints. If the number of
constraints is lower than the degrees of freedom, there are zero frequency
vibrational modes called floppy \cite{Thorpe1}, and the resulting network is
under-constrained. A transition occurs when the lattice becomes rigid, and
at the corresponding chemical composition, the glass is easy to form. Many
features of this transition have been experimentally observed \cite
{Boolchand1}\cite{Boolchand2}. \ Also, one of the authors proposed that
rigidity can be related with the statistics of the phase space energy
landscape \cite{Naumis}, since the number of floppy modes is equal to the
number of different configurations of the system with nearly equal minimal
energies \cite{Naumis}, and thus is a way to evaluate the function that
gives the number of minima energy basins \cite{Debenedetti}.

In a recent paper, Thorpe {\it et. al.} \cite{Thorpe2}, remarked that in
real glasses, even though formed at relative high temperatures, where the
entropic effects are dominant, it is not correct to fully ignore energetic
contributions which can favor particular structural arrangements over
others, ({\it e.g.} in a binary system chemical aggregation between unlike
particles favor local chemical aggregation). One interesting question that
they address is how the structure itself can incorporate non-random features
in order to minimize the free energy at the temperature of formation. They
answer this question by proposing that the structure can self-organize
avoiding stress in the random formed network \cite{Thorpe2}. In the
literature exists experimental evidence for self-organization in glasses 
\cite{Boolchand}, this evidence has been associated with the intermediate
phase proposed by Phillips \cite{Phillips2}. In a previous work \cite{huerta}%
, we observed that in a model of an associative fluid (the Cummings-Stell
model), some thermodynamics features can be associated with a rigidity
transition, and in particular, it was shown that a glass transition occurs
very near to the RT. Also, we showed that using the MC step as a time
parameter in a NPT ensemble, we were able to control the cooling rate of a
liquid melt in a qualitative way. In this work, we go further by looking at
the self-organization properties in the Cummings-Stell model (CS), using a
Monte-Carlo (MC) computer simulation in a grand canonical (GC) ensemble.
Compared with the NPT ensemble, the GC ensemble has the advantage of
reaching equilibrium faster \cite{mezard}, given the opportunity to visit a
wider range of equilibrium an non-equilibrium phases. In order to talk about
the thermodynamic properties of these phases, we basis our assumptions in
the fact that their life time is larger than the observation time (averaged
time) \cite{debenebook}. This time is also larger than the molecular
relaxation time, which we can adjust by tuning the MC steps of single
particle movements, and the MC steps of the formed clusters movements. As a
consequence, the slower a liquid is cooled, the longer the time available
for configuration sampling at each temperature, and hence the access to the
homogeneous nucleation which leads to crystallization. We point that this
nucleation produces stress in the obtained structure, as a counting of
floppy modes reveals. As a counter part, the faster the liquid is cooled,
there is less time available for homogeneous nucleation and hence less
stress is produced in the structure, inducing a local self-organization.
This framework allow us to address the question of what are the structural
and thermodynamic properties of a self-organized structures and how they
behave as the system is cooled. With this in mind, we perform MC simulations
where configurations that produce stress in the system are rejected, in a
similar way to that proposed by Thorpe {\it et. al}. \cite{Thorpe2} for
studying rigidity self-organization. As we will see, our results are in
agreement with ref. \cite{Thorpe2}, \ since the avoidance of stress, bias
the system to a glass state. The layout of this work is as follows, in
section II we introduce the model to be used, in section III a method for
indirect controlling of the various relaxation times is introduced, and in
section IV we discuss the effect of self-organization. Finally, in section V
the conclusions of the work are given.

\section{Model}

We choose a simple model of an associative fluid: the Cummings-Stell model
(CSM) of a two component system ($A$ and $B$) of associating disks in 2D,
both of the same size \cite{huerta}. The particles interact via a potential
permitting core interpenetration of the $A$ and $B$ monomer discs, so that
the bond length $L$ is less than the core diameter $\sigma $. Without loss
of generality we assume $\sigma =1$. The interactions are given as follows:

\begin{eqnarray*}
U_{ij}(r) &=&U_{ij}^{hd}(r)+(1-\delta _{ij})U_{as}(r), \\
U_{AA}^{hd}(r) &=&U_{BB}^{hd}(r)=\left\{ 
\begin{array}{c}
\infty \qquad r<1, \\ 
0\qquad r>1,
\end{array}
\right. \\
U_{AB}^{hd}(r) &=&U_{BA}^{hd}(r)=\left\{ 
\begin{array}{l}
\infty \qquad r<L-0.5w, \\ 
D\qquad L-0.5w<r<1, \\ 
0\qquad r>1,
\end{array}
\right. \\
U_{as}(r) &=&\left\{ 
\begin{array}{l}
0\qquad r<L-0.5w, \\ 
-\varepsilon _{as}-D\qquad L-0.5w<r<L+0.5w, \\ 
0\qquad r>L+0.5w,
\end{array}
\right.
\end{eqnarray*}
where $i$ and $j$ stand for the species of the particles and take values $A$
and $B$. $r$ is the separation between centers, $L$ is the bonding distance
and $w$ is the width of the attractive intracore square well (fig. 1). The
model allows the formation of dimer species for small values of the bonding
length parameter, the formation of chains, if the bonding length is slightly
larger, and also vulcanization with fixed maximum coordination number for
different bonding length values close to the diameter of particles, as shown
in fig. 1. In order to be able to fix a maximum coordination number in each
simulation , we take $D\rightarrow \infty $ \ as was done before in other
works \cite{huerta}, \cite{huerta0}, \cite{stell}. This choice has the
effect that unlike particles avoid bond-lengths between $L+0.5w$ and $1,$
and thus coordinations higher than a desired maximum are not allowed.
Numerically, this condition means that in the MC simulations, we never
consider bond distances in the previous range. \ The corresponding values
for each maximum coordination number are given in the following table,

\bigskip

\begin{tabular}{|c|c|c|}
\hline
$r_{\max }$ & $L$ & $w$ \\ \hline
cpx $3$ & $0.65$ & $0.1$ \\ 
cpx $4$ & $0.78$ & $0.1$ \\ 
cpx $5$ & $0.91$ & $0.1$ \\ \hline
\end{tabular}

Table 1. Parameters of the CS model that fix the maximum coordination of the
particles as used in this work. The notation cpx $r$, means complex of
particles with maximum coordination $r_{\max }$.

\section{Glass transition by controlling relaxation times}

We start by pointing out that a supercooled liquid phase is metastable with
respect to the crystalline state, and this supercooling can be achieved if
nucleation is inhibited during cooling \cite{debenebook} . One way of
inhibit nucleation is by performing a rapid quench of the liquid; in these
terms, two characteristic times $\tau _{1}$ (the time required for
crystallization) and $\tau _{2}$ (the time corresponding to molecular
relaxation) compete between crystallization and vitrification \cite
{debenebook}. In this work, we simulate this effect in two different ways,
which we will show that at the end turn out to be very similar: one is to
use the MC steps as a time parameter in the GC ensemble \cite{Newman}, where
we tune the ratio between $\tau _{1}$ and $\tau _{2}$ in an indirect way, by
controlling the ratio between steps of particle and cluster rearrangements,
since the first is the most important factor for molecular relaxation, while
the second optimize crystallization. The second way which we observed that
leads to supercooling is the self-organization of rigidity, as we will see
in the next section.

To implement supercooling using the MC steps by indirect control of the
relaxation times, we use a GC Metropolis Monte-Carlo method \cite{Allen}.
The procedure has two nested loops. In the inner one, the particles are
moved inside the volume, and an interchange of particles with the particle
reservoir is allowed. This loop is performed $N_{m}$ times. The particle
movements inside the volume allow to rearrange the structure, and thus this
is related with the molecular relaxation of the structure ($\tau _{2}$). In
the outer loop, cluster rearrangements and the average of the
thermodynamical quantities are performed, each time that $N_{m}$ cycles of
the inner loop are finished. The external loop is related with
crystallization, since cluster movements promote the growth of bigger
clusters. Is clear that if $N_{m}$ is high enough, the probability of having
local cluster nucleation is high, and thus cluster movements allow to form a
crystal by successive aggregation of small clusters. When $N_{m}$ is small,
the local configurations are not in equilibrium, and hence the cluster
movements promote the generation of a random network.

In fig. 2, we show the results of the inverse density ($\rho ^{-1}$) against
reduced temperature ($T^{\ast }=(\beta \varepsilon _{as})^{-1}$) for several 
$N_{m}$ cycles which simulate different $\tau _{2}$ times. In this figure,
we fixed the parameters of the CSM that allows as maximum coordination four
(cpx4), $L=0.78,w=0.1$, restricted to equimolar concentration $\beta \mu
_{A}=\beta \mu _{B}=-0.4$. Finally, after each $N_{m}$ steps of the inner
cycle, we allow the possibility of one cycle cluster rearrangement. In order
to simulate the same cooling rate with several $\tau _{2}$ times, we fix the
outer cycle for $100$ steps during the quasi-equilibration run, and $2000$
times for a productive run. In such a way, we averaged over the same number
of configurations for each different $N_{m}$. As can be seen, for $%
N_{m}=1500 $ a crystallization is observed, while for the other values, a
glass transition is obtained, as is revealed by the figure and by a direct
inspection of the resulting structures. In all the procedure, MC steps are
controlled to have an acceptance ratio between $20-30$\%.

An interesting observation, is that the fraction of particles with maximum
coordination $\chi _{4}$, depends strongly on $N_{m}$. In table 2, we show
this fraction against temperature for the same simulations presented in fig.
2. When the crystal is formed, $\chi _{4}$ is nearly one, while for the
supercooled liquid at the same temperature, $\chi _{4}$ remains at the same
order of magnitude. We can understand this effect as follows: if full
thermal equilibrium of the system is not allowed, \ is not possible to
access the global minimum of the energy potential \cite{Corti}, and hence
the nucleation is prohibited, with the consequence that the supercooled
liquid is structurally arrested at a finite temperature and restricted to
explore the configurational space correspondent to a single basin in the
energy landscape. In the next section, we use this idea to relate this
observation with the self-organization of rigidity.

\bigskip

\bigskip 
\begin{tabular}{|c|c|c|c|c|}
\hline
$T^{\ast }$ & $N_{m}=12$ & $N_{m}=40$ & $N_{m}=150$ & $N_{m}=1500$ \\ \hline
0.40 & 0.0011 & 0.00113 & 0.0025 & 0.0019 \\ 
0.38 & 0.0046 & 0.00103 & 0.0053 & 0.0040 \\ 
0.36 & 0.0022 & 0.00404 & 0.0043 & 0.0079 \\ 
0.34 & 0.0007 & 0.01287 & 0.0152 & 0.0188 \\ 
0.32 & 0.0077 & 0.01677 & 0.0495 & 0.0591 \\ 
0.30 & 0.0108 & 0.04376 & 0.1353 & 0.9395 \\ 
0.28 & 0.0258 & 0.05391 & 0.1538 & 0.9786 \\ 
0.26 & 0.0366 & 0.05138 & 0.2229 & 0.9666 \\ 
0.24 & 0.0725 & 0.07417 & 0.3100 & 0.9487 \\ 
0.22 & 0.1057 & 0.08355 & 0.3292 & 0.9622 \\ 
0.20 & 0.1116 & 0.07762 & 0.3496 & 0.9617 \\ \hline
\end{tabular}

Table 2. Fraction of maximum coordinated particles as a function of the
reduced temperature ($T^{\ast }$) and molecular relaxation time, controlled
by the parameter $N_{m}$.

\section{\protect\bigskip Glass transition by self-organization of rigidity}

The fact that the maximum coordination is not achieved for most of the
particles in the supercooled liquid, means that inhibit nucleation is a
natural way of inhibit crystallization, as was discussed in the
introduction. This simple idea can be put in contact with the rigidity ideas
of Phillips \cite{Phillips2} and Thorpe \cite{Thorpe1}. As we mention
before, in this theory, the ability for making a glass is optimized when the
number of freedom degrees, in this case $2N,$ where $N$ is the number of
particles, is equal to the number of mechanical constraints ($N_{c}$) that
are given by the bond length and angles between bonds.

$(2N-N_{c})/2N$ gives the fraction of cyclic variables of the Hamiltonian,
and also corresponds to the number of vibrational modes with zero frequency (%
$f$), called floppy modes,\ with respect to the total number of vibrational
modes. The counting of floppy modes in a mean-field, known as Maxwell
counting, goes as follows: since each of the $r$ bonds in a site of
coordination $r$ is shared by two sites, there are $r/2$ constraints due to
distance fixing between neighbours. If the angles are also rigid, in $2D$
there are $(r-1)$ constraints, to give, 
\[
f=\frac{2N-N_{c}}{2N}=1-\frac{<r>}{4}-\sum_{r}(r-1)x_{r} 
\]
where the last term corresponds to the angular constraints, $\chi _{r}$ is
the fraction of particles with coordination $r$, and $<r>$ is the average
coordination number, defined as, 
\[
<r>=\sum_{r}rx_{r} 
\]
A rigidity transition occurs when $f=0$ and the system pass from a floppy
network to rigid one. If $f$ is a negative number, {\it i.e.}, if there are
more constraints than degrees of freedom, the lattice is overconstrained and 
$f$ is the number of stressed bonds. In $2D$, the rigidity transition leads
to the critical value $<r>=2.0$ if all angular constraints are considered,
and $<r>=4.0$ if the angular restoring forces are not strong.

Within the Cummings-Stell model, rigidity comes from the association of
particles: each bond generates a constraint, and the angular constraints are
only produced by geometrical hindrance, {\it i.e.}, the angles between
particles can change without a cost in energy, but within certain limits
imposed by the restriction of the hard-core interaction between like
particles, as shown in fig. 3. For maximum coordination four, this means
that only sites with coordination four have a contribution to angular
constraints. It is true that sites with coordination two and three in
principle should provide extra angular constraints, since the hard core
interaction gives a minimum angle between particles. However, the angles are
not fixed and they have a wide region to allow particle movements, and thus
do not contribute to the restriction counting.

Taking into account the geometrical hindrance of the model, the number of
floppy modes is now given by,

\[
f=1-\frac{<r>}{4}-\sum_{r}\delta _{rr_{\max }}(r-1)x_{r} 
\]

where $r_{\max }$ is the maximum allowed coordination, and $\delta
_{rr_{\max }}$ is a Kronecker delta. From here, is clear that when in a
cluster we have a site with maximum coordination, rigidity raises since the
delta function is different from zero and more constraints are added that
over-constraint the cluster. Thus, sites with maximum coordination nucleate
rigidity and produce stress in the lattice. For example, in a crystal with
coordination four, $\chi _{4}=1$ and $f=-3/2$, which means that the lattice
is over-constrained. According to Phillips, when $f=0$, it is easy to form a
glass, since the material is neither over-constrained (that produce
explosive exothermic crystallization due to strain energy\cite{Phillips3})
nor under-constrained (leading to the formation of a molecular crystal), the
system is trapped in a configurational limbo \cite{Phillips3}, where
fluctuations do not provide a pathway to the crystalline phase.

Furthermore, in the last section we have showed that the fraction of
particles with maximum coordination is in close connection with the
molecular relaxation time, which in other words means that to form a glass,
nucleation of stress must be prevented. From our previous results, we can
observe that the probability of formation of a nucleated structure is small
due to the high molecular relaxation time. We decided to follow these ideas
by proceeding in the opposite way than in the last section, {\it i.e.}, we
inhibit rigidity nucleation by rejecting configurations with maximum
coordination and then we see if we are able to bias the system to a glassy
state, in such a way that we simulate long molecular relaxation times.
Observe that rejecting configurations that produce stress is the same
process of self-organization that was considered by Thorpe {\it et. al.} in
order to form stress-free lattices \cite{Thorpe2}. In that sense, we look if
self-organization of rigidity is able to produce a glass transition. This
kind of simulation is usually called biased Monte-Carlo \cite{Frenkel}.

To study the effect of self-organization, we made the same MC procedure
described in the previous section, but with $N_{m}$ fixed to the value that
gives crystallization ($N_{m}=1500$). The only difference with the previous
case is that now we reject particle movements that produce a site with
maximum coordination.

In figure 4, we present the behaviour of the inverse of the density $(\rho
^{-1})$ as a function of the reduced temperature $T^{\ast }=(\beta
\varepsilon _{as})^{-1}$, with the condition that allows maximum four
neighbours (cpx4), restricted to equimolar concentration $\beta \mu
_{A}=\beta \mu _{B}=-0.4$, (open squares). As the temperature is slowed
down, we can observe a continuous decrease in $\rho ^{-1}$. However, for
reduced temperatures lower that $0.30$, a jump in $\rho ^{-1}$ is observed
when all the configurations are allowed. This jump corresponds to the
crystalline like phase transition, as can be argued by the shape of the
transition, from an inspection of the configuration obtained, and by the
radial distribution function. Due to the fact that it is possible to keep
the system without stress, we develop the same simulation as before but
rejecting in the simulation every configuration that contains a particle
with coordination four. The results are presented in fig.4 with dashed
squares. In that case, the system remains as a supercooled fluid. Moreover,
the system can not form a crystal structure as occurs in the simulation in
which we allow stress, and thus do not present a usual phase transition,
instead a glass like transition is observed. These results shows that
self-organization of rigidity is able to produce a glass.

An important remark is that avoiding configurations with maximum
coordination is not equivalent to consider a CSM without self-organization
but with a lower maximum coordination. For example, in fig. 4 we plot the
results of a simulation without rejection for a CSM that allows maximum
coordination three (cpx3). As can be seen, the model also presents
crystallization.

In figure 5, we present the results for the same kind of simulation but for
a system that allows maximum coordination three (cpx3). As can be seen, the
rejection of stressed configurations also leads to a glass like transition.

Now we turn our interest to the condition of maximum coordination five
(cpx5), as show in fig. 6. As can be seen, in this case the glass transition
is produced even when the stressed configurations are rejected. This fact
can be understood in terms of \ rigidity in the following way: when $r_{\max
}=5$, it is impossible to have a crystal due to geometric frustration at
equimolar condition $\beta \mu _{A}=\beta \mu _{B}=-0.4$, and $\chi _{5}\ll
1.$ Since the rigidity transition without angular restrictions occurs only
when $<r>=4,$ most of the configurations do not produce stress and the
system behaves freely (i.e. we do not need to reject any configuration) as a
self-organized system. In this sense, geometric frustration induce
self-organization of the system. As a corroboration of this fact, in fig.7
we show the number of floppy modes as a function of the average coordination
number, using the Maxwell counting. We remark that each coordination
corresponds to a certain temperature of the simulation. For example, at high
temperatures, all the models with different maximum coordination fall in the
same line, since in the liquid the probability of nucleation is very low.
However, for the case of cpx5, cpx3 and cpx4 without stress, all the
simulations fall again in the same line even for low temperatures, since the
self-organization means that the clusters grow without angular constraints
(stress free). When this line is extrapolated to $f=0$, we obtain $<r>=4$,
which is the value for a rigidity transition without angular constraints. If
angular constraints are allowed, the simulations for low temperatures falls
outside the line determined by self-organization, and the rigidity
transition occurs at lower values of $<r>.$ Finally, we can compare these
results to the floppy mode counting made for the glass transition using the
method of tuning the different relaxation times. In fig. 8, we present the
number of floppy modes as a function of $<r>$. As can be seen, when $N_{m}$
is high, there is a transition of rigidity due to nucleation, while for low $%
N_{m}$ the system tends to stay in the line of self-organization.

\section{\protect\bigskip Conclusions}

In this work, we have explored the connection between self-organization of
rigidity, and the supercooling of a liquid to form a glass. By considering
an associative fluid model, we showed that the competition between two
different characteristic times, molecular relaxation and crystallization
times, can be modelled using a MC simulation, where the number of cycles
between particle and cluster moves is controlled. The results of these
simulations, suggested that nucleation is avoided during supercooling and
produce self-organization in the sense of the rigidity theory. This idea was
also tested by making MC simulations but avoiding stressed configurations.
As a result, we were able to produce glass transitions using
self-organization. In a model (cpx5) with geometrical frustration, this
self-organization is provided by geometry, and thus glass transition occurs
without rejecting configurations. All of the results of this article are in
agreement with the Phillip%
\'{}%
s idea that glass transition is related with rigidity due to the lack of a
pathway to crystallization \cite{Phillips3}. \ Many of these facts, can also
be studied form an energy landscape point of view, as we will show in future
works.

{\bf Acknowledgments.} This work was supported by DGAPA UNAM project
IN108199, and the supercomputer facilities of the DGSCA-UNAM. A.H. thanks
the economical and credit supports given by CONACYT project GO010-E and ref.
167165

.

1.-Cummings-Stell model.

2.- Inverse of the density ($\rho ^{-1}$) as a function of the scaled
temperature ($T^{\ast }$) for different values of $N_{m}$.

3.-Counting of angular constraints in the Cummings-Stell model. A cluster of
two particles has no angular constraints, since one of the particles can
rotate $360%
{{}^\circ}%
$ around the other, while a in a cluster with coordination four the angle
between particles is fixed, which leads to $3$ angular constraints.

4.-The same as fig. 2 for the case of stress free nucleation for a system
with maximum allowed coordination four (cpx4), with and without stress
(squares and dashed squares). We include a simulation for maximum
coordination three (cpx3) with stress.

5.- Inverse of the density ($\rho ^{-1}$) as a function of the scaled
temperature ($T^{\ast }$) for maximum coordination three with and without
stress (triangles and dashed triangles).

6.-Inverse of the density ($\rho ^{-1}$) as a function of the scaled
temperature ($T^{\ast }$) for maximum coordination five (cpx5) with and
without stress (pentagons and dashed pentagons). For comparison purposes, we
include a simulation of cpx4 without rejecting configurations (squares).

7.-Number of floppy modes as a function of the coordination number \ ($<r>$)
for each of the models with different maximum coordination number.

8.-Number of floppy modes as a function of $<r>$ for several $N_{m}$
relaxation times without rejecting any configuration.

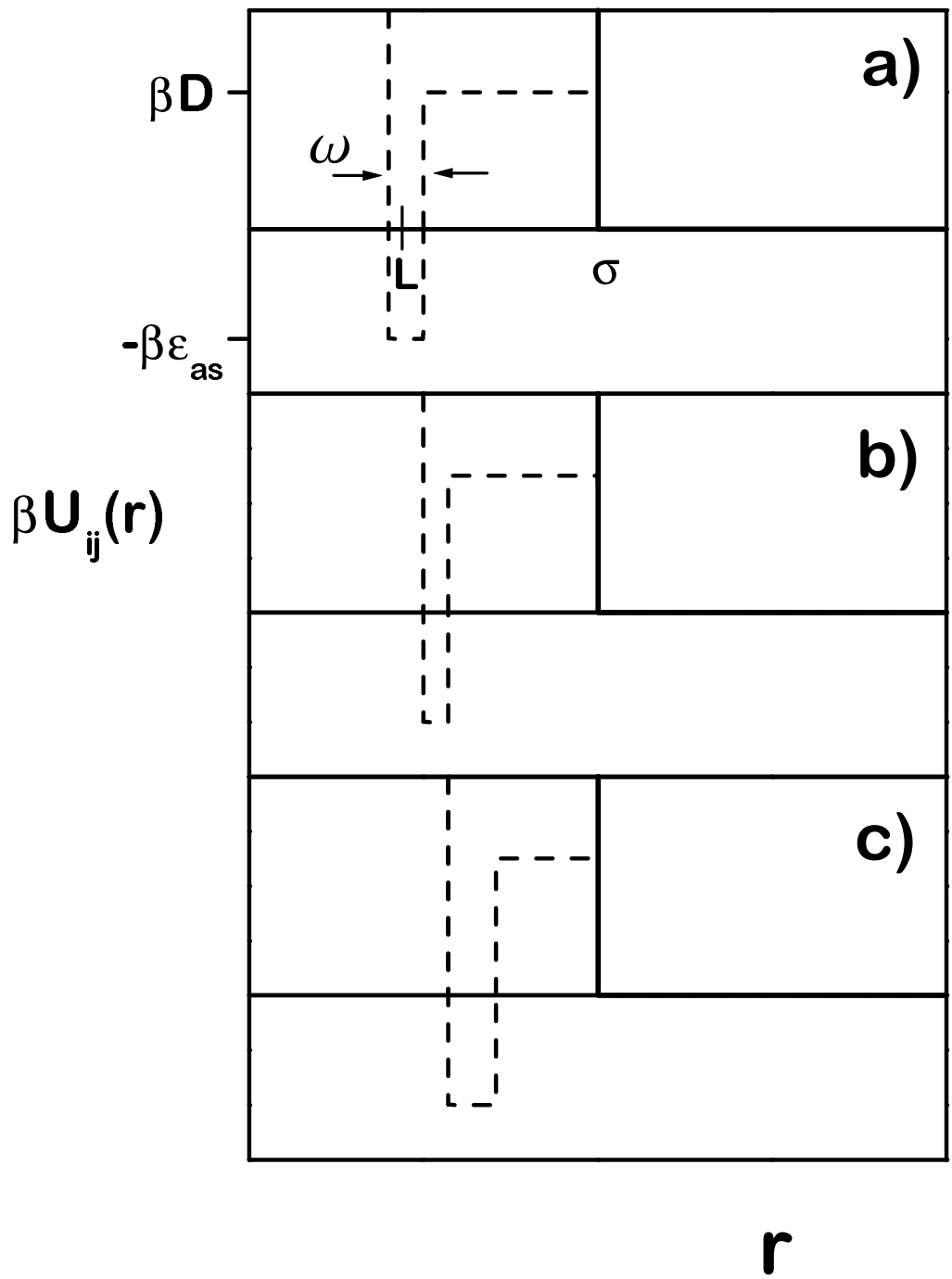
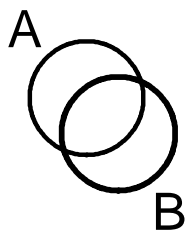
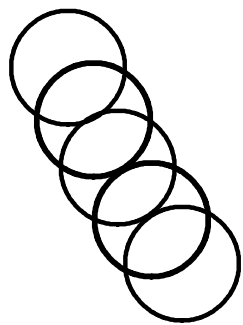
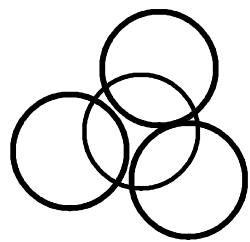
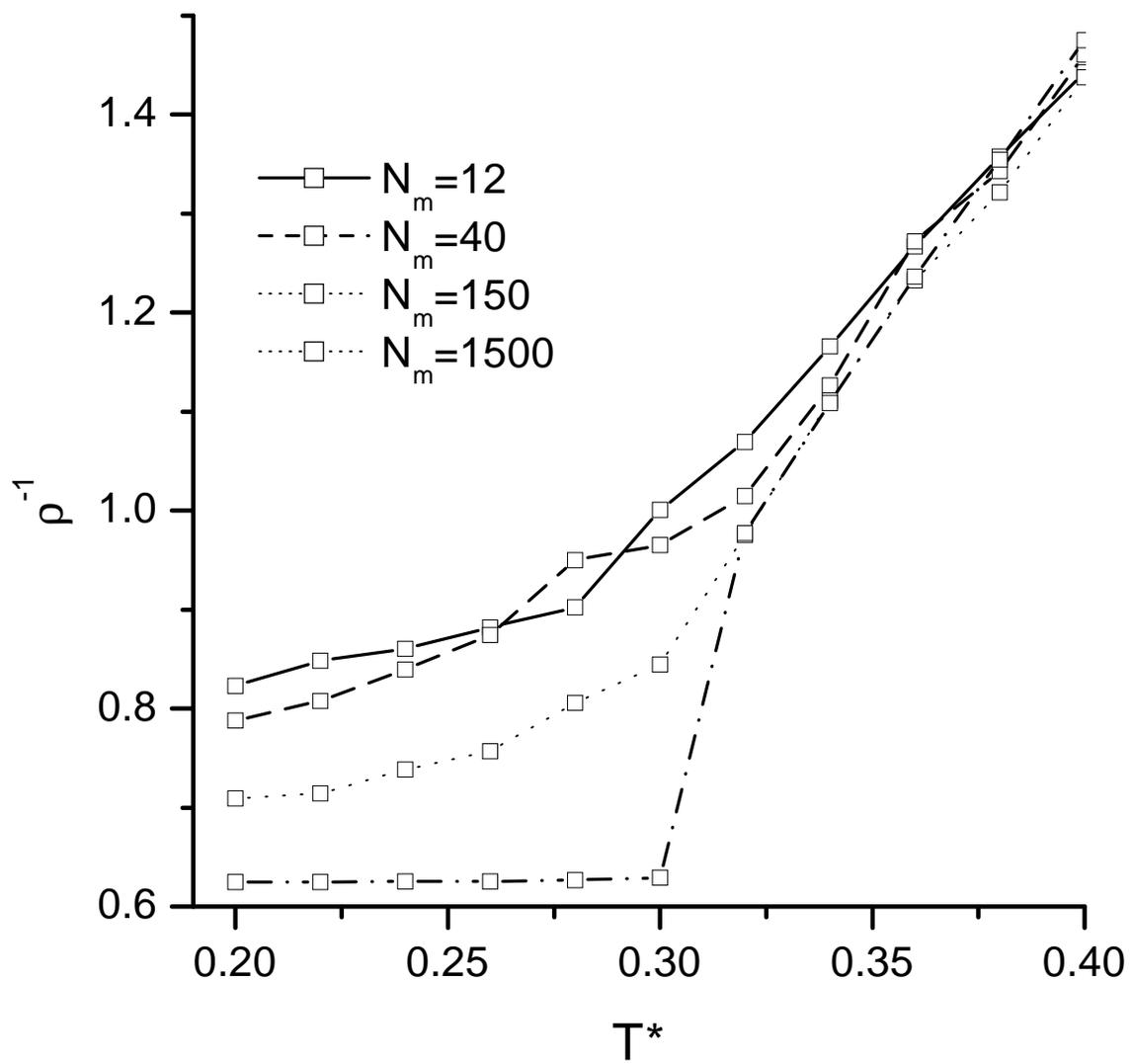
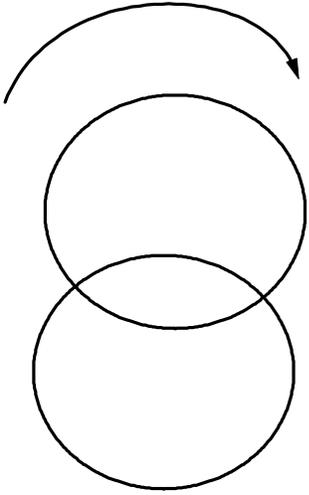
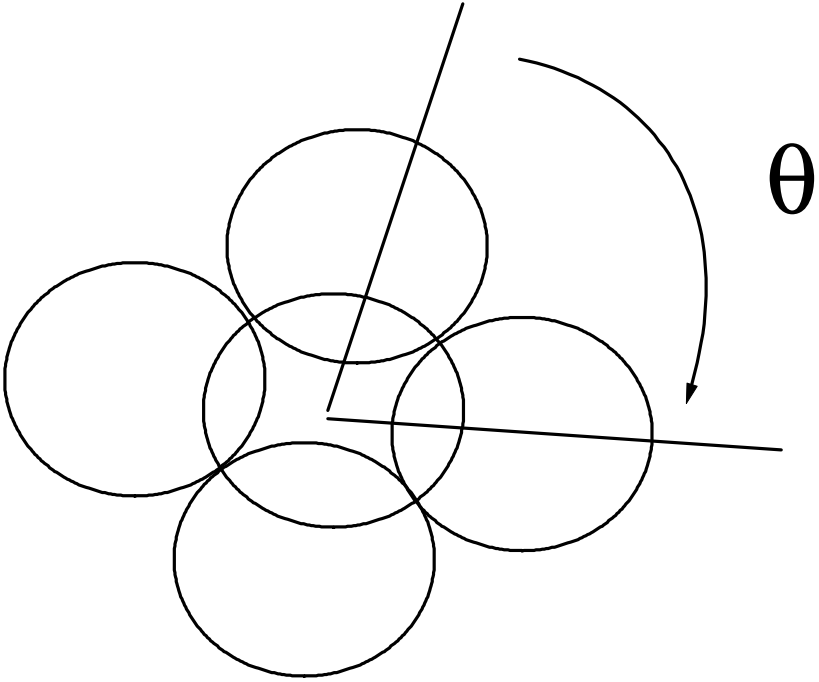
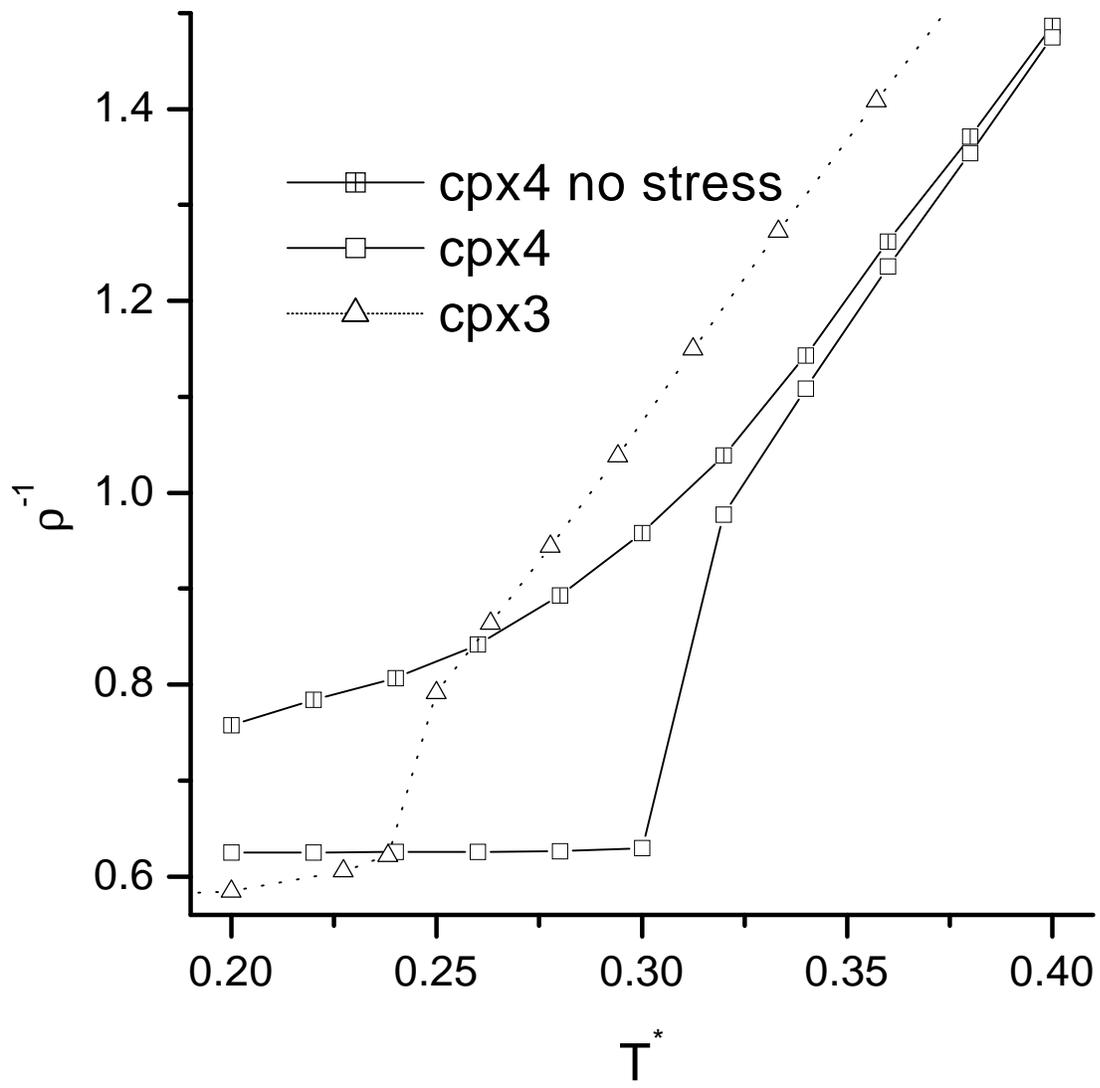
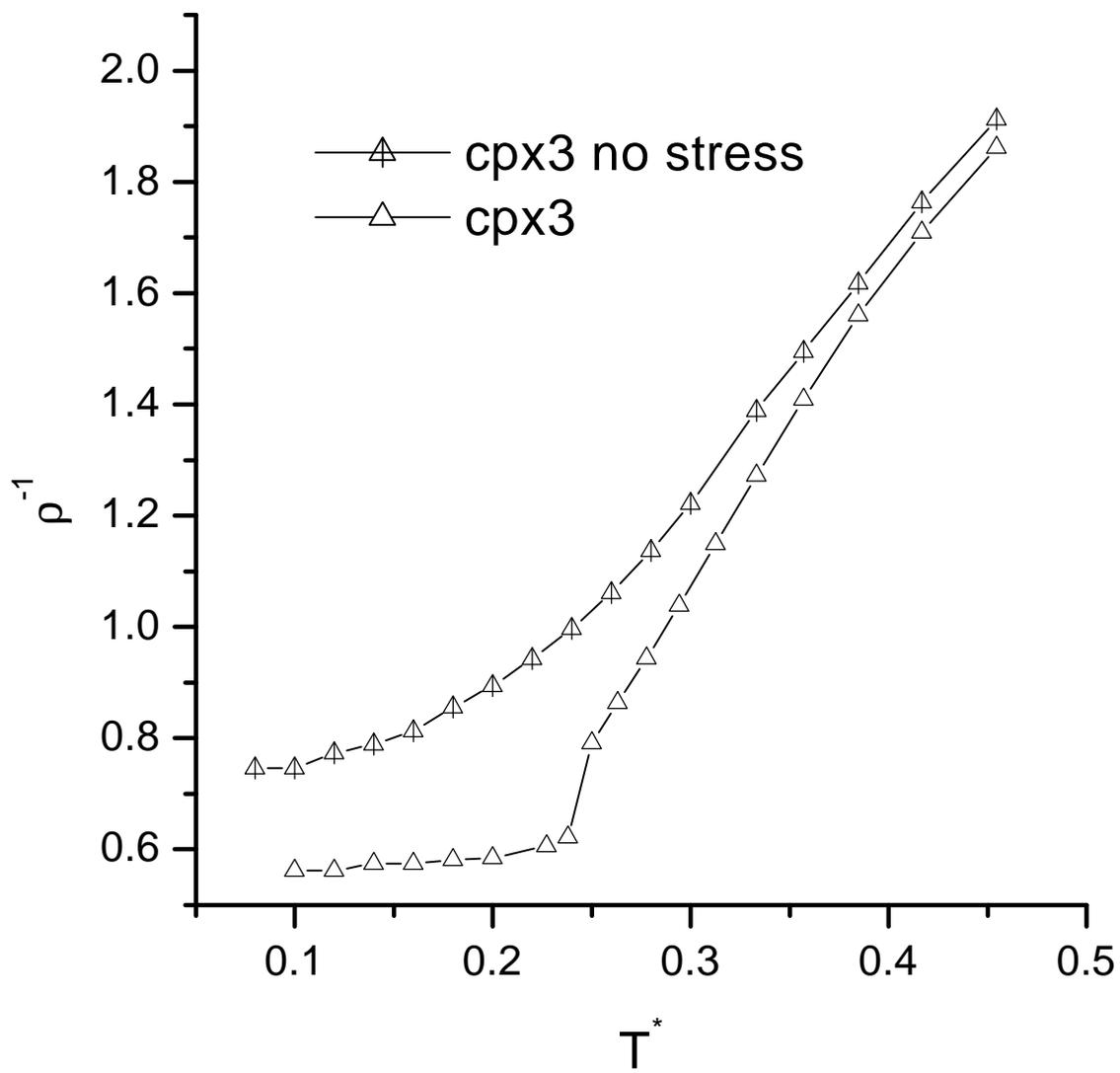
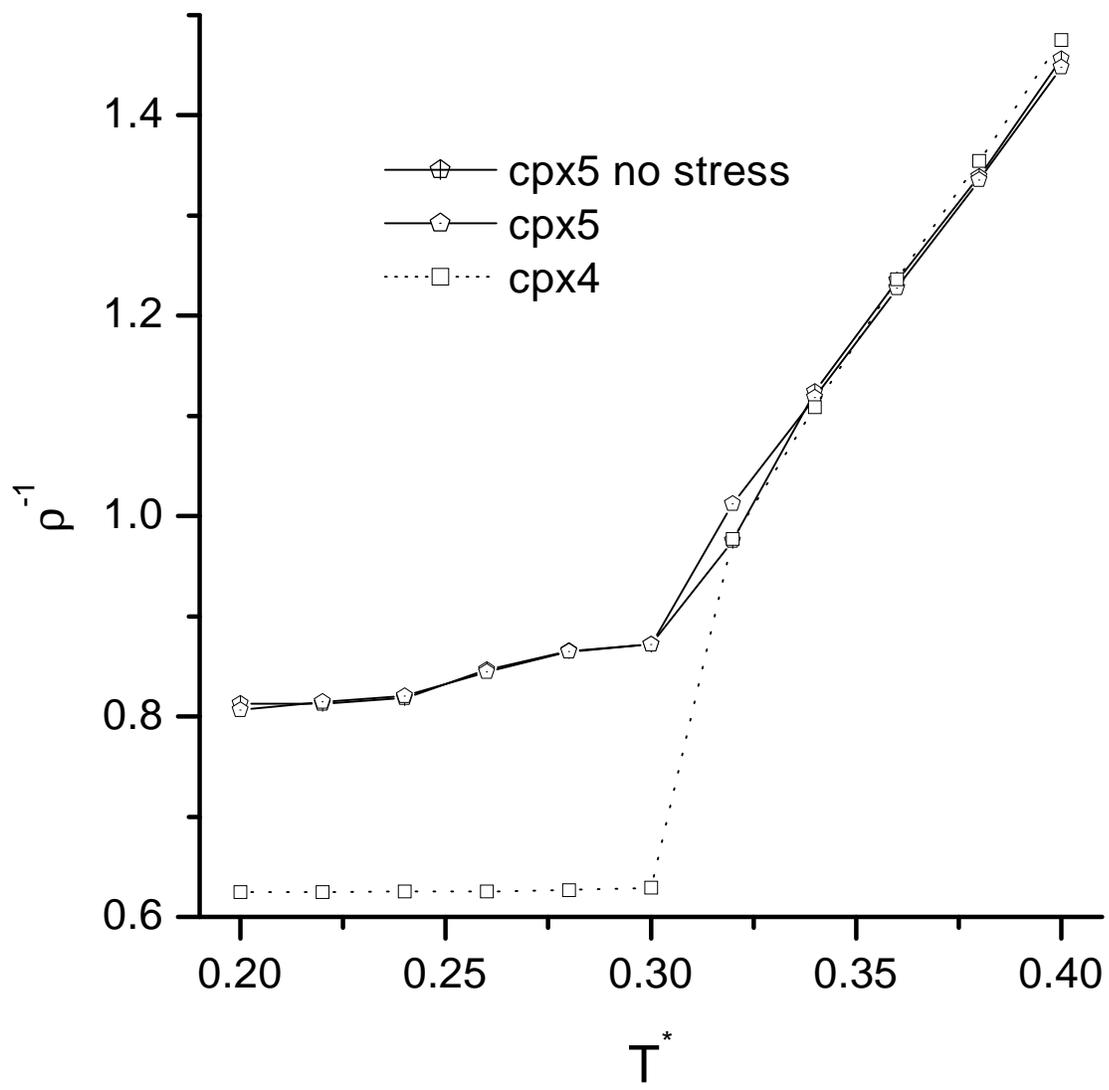
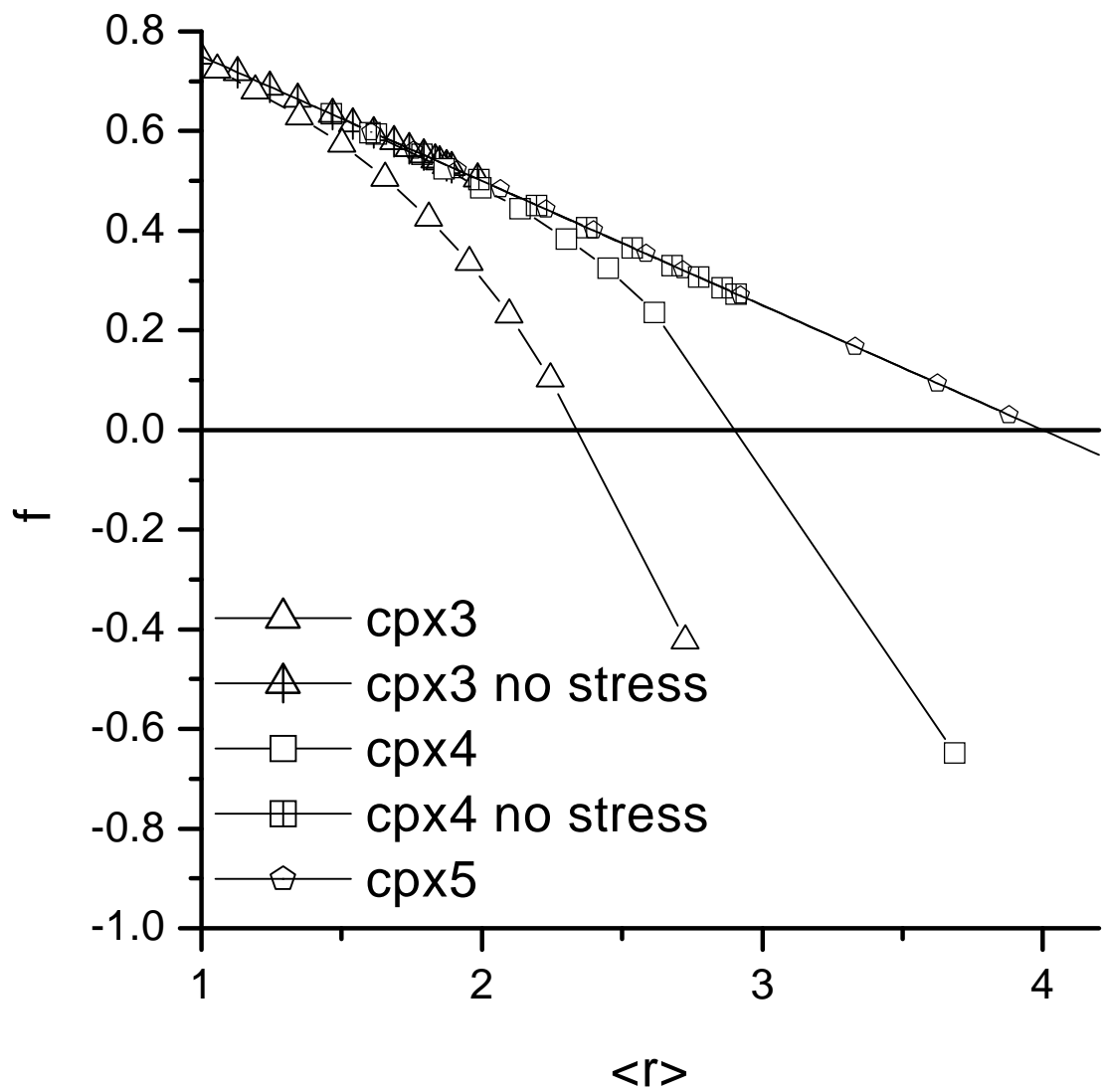
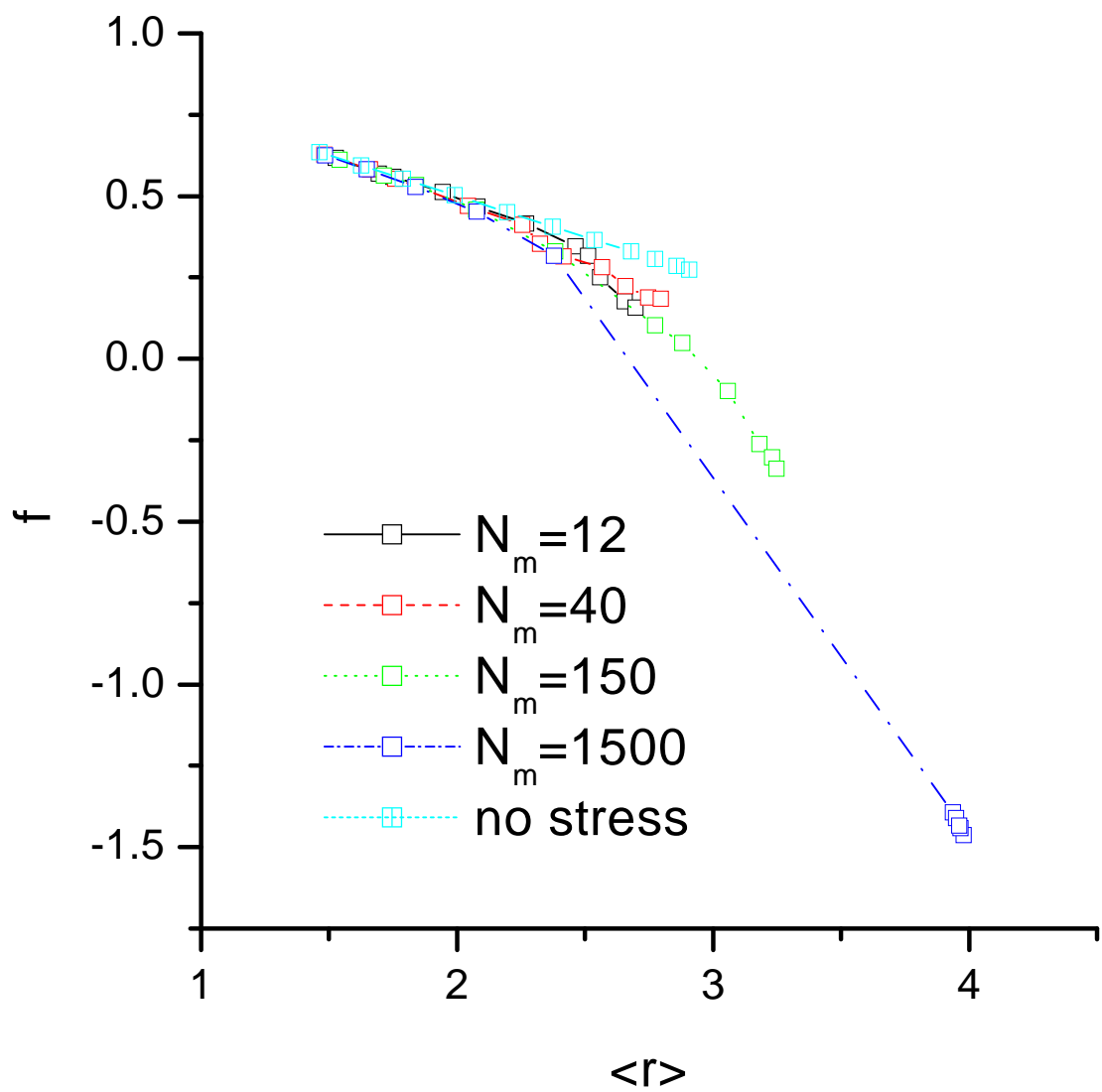
\end{document}